\documentclass[aps,pre,preprint,groupedaddress]{revtex4}

\usepackage{graphicx}
\usepackage{latexsym}
\usepackage{amsmath,amssymb}

\begin{document}

%\preprint{}

\title{
Multicriticality of the $(2+1)$-dimensional gonihedric model:
A realization of the $(d,m)=(3,2)$ Lifshitz point
}

\author{Yoshihiro Nishiyama}
%\email[]{Your e-mail address}
%\homepage[]{Your web page}
%\thanks{}
%\altaffiliation{}
\affiliation{Department of Physics, Faculty of Science,
Okayama University, Okayama 700-8530, Japan}

\date{\today}

\begin{abstract}
Multicriticality of the gonihedric model in $2+1$ dimensions is investigated numerically.
The gonihedric model is a fully frustrated Ising magnet with
the finely tuned plaquette-type (four-body and plaquette-diagonal) interactions,
which cancel out the domain-wall surface tension.
Because the quantum-mechanical fluctuation along the imaginary-time direction
is simply ferromagnetic,
the criticality of
the $(2+1)$-dimensional gonihedric model should be an anisotropic one;
that is, the respective critical indices of
real-space  ($\perp$) and imaginary-time ($\parallel$) sectors
do not coincide.
Extending the parameter space to control the domain-wall surface tension,
we analyze the criticality in terms of the crossover (multicritical)
scaling theory.
By means of the numerical diagonalization for the clusters
with $N \le 28$ spins,
we obtained the correlation-length critical indices
$(\nu_\perp,\nu_\parallel)=(0.45(10),1.04(27))$,
and the crossover exponent $\phi=0.7(2)$.
Our results are comparable to $(\nu_{\perp},\nu_{\parallel})=(0.482,1.230)$, and 
$\phi=0.688$
obtained by Diehl and Shpot 
for the $(d,m)=(3,2)$ Lifshitz point
with the $\epsilon$-expansion method up to O$(\epsilon^2)$.
\end{abstract}

% insert suggested PACS numbers in braces on next line
% insert suggested keywords - APS authors don't need to do this
%\keywords{}

%\maketitle must follow title, authors, abstract, \pacs, and \keywords
\maketitle

\section{\label{section1}Introduction}

Recently, a thorough investigation of the Lifshitz point
was made by Diehl and Shpot with the $\epsilon$-expansion
method up to O$(\epsilon^2)$
\cite{Diehl00,Shpot01}; see also Refs. 
\cite{Diehl03,Leite03a,Leite03b,Hornreich75,Mergulhao98,Mergulhao99}.
The field theory for the Lifshitz point 
has an anisotropic dispersion like 
$\omega({\bf k})=\frac{1}{2}\sum_{i=1}^{m}k_i^4 +\frac{1}{2}\sum_{i=m+1}^{d}k_i^2$,
preventing us from going beyond order O$(\epsilon^2)$.
Reflecting this anisotropy,
the critical indices 
within the subspaces, $i=1,2,\dots,m$ ($\perp$) and $i=m+1,m+2,\dots,d$ ($\parallel$),
are no longer identical.
In Ref. \cite{Shpot01}, the critical indices within each subspace are tabulated
systematically
for the generic values of $(d,m)$.

Such an anisotropic criticality 
is realized by 
the $d$-dimensional Ising model
fully-frustrated within the $m$-dimensional subspace.
The problem is that
a naive computer simulation for the equilateral cluster
does not yield adequate finite-size scaling.
Rather,
one has to adjust the shape of the cluster
(that is,
the system sizes of each subspace  $L_{\parallel,\perp}$)
so as to fix the following scaled ratio to a constant value;
\begin{equation}
\label{constraint}
L_\perp^z /  L_\parallel =const  .
\end{equation}
Here, the index $z$ denotes the dynamical critical exponent,
which characterizes the anisotropy.
The significant point is that
the exponent $z$ itself is an unknown parameter,
and it has to be determined through some preliminary analyses.
After that, one is able to perform large-scale simulations.
So far, the case of $(d,m)=(3,1)$, 
namely, the axial-next-nearest-neighbor-Ising model,
has been studied extensively by means of the Monte Carlo method
\cite{Selke78,Kaski85,Pleimling01}.
The simulation results are in agreement with
the above-mentioned field-theoretical considerations
as well as
the series-expansion results \cite{Oitmaa85,Mo91}.

\begin{figure}
\includegraphics[width=100mm]{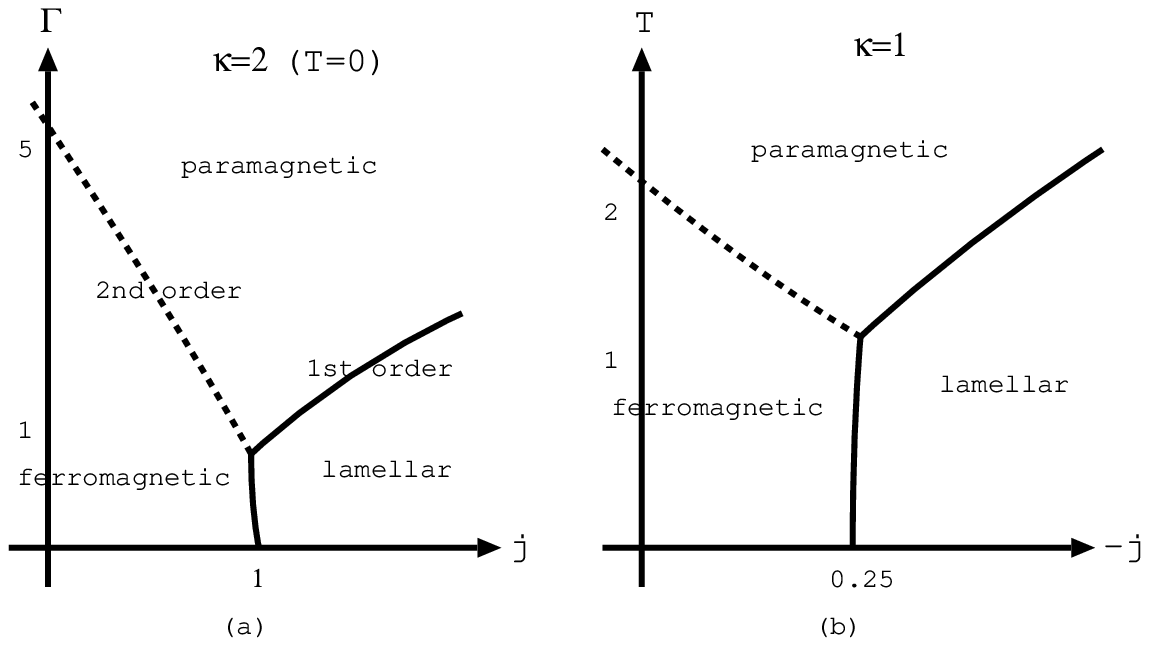}%
\caption{
\label{figure1}
(a)
A schematic drawing of the ground-state phase diagram of the $(2+1)$-dimensional gonihedric model,
Eq. (\ref{Hamiltonian}), with the self-avoidance parameter $\kappa=2$.     %%%  4$.
We aim to investigate the multicritical point at $j=1$.
(b) As a comparison, we present the phase diagram of
the $d=3$ (classical) gonihedric model, Eq. (\ref{Hamiltonian_classical}),
with $\kappa=1$ \cite{Cirillo97b,Nishiyama04};
here, the parameter $T$ denotes the temperature.
The phase diagram is essentially the same as that of
the quantum-mechanical model;
the discrepancy $j \leftrightarrow -j$ is due to the
difference of 
parameterization.
}
\end{figure}

In this paper, we consider
the case of $(d,m)=(3,2)$.
For that purpose, we investigate
the ground-state phase transition of the gonihedric model in 
$2+1$ dimensions.
The gonihedric model is a fully frustrated Ising magnet
with the finely tuned plaquette-type (four-body and plaquette-diagonal) interactions,
for which the domain-wall surface tension vanishes;
so far, the classical version has been studied in detail
\cite{Savvidy94,Ambartzumian92,Cirillo97b,Koutsoumbas02,Espriu97}.
Making a contrast to the frustrated magnetism within the real space ($\perp$),
the quantum fluctuation along the imaginary-time direction ($\parallel$) is simply 
ferromagnetic,
and the ground-state criticality should be an anisotropic one.
In Fig. \ref{figure1} (a),
we present a schematic phase diagram 
of the $(2+1)$-dimensional gonihedric model
subjected to the transverse
magnetic field $\Gamma$ and the frustration $j$;
we explain the details in Sec. \ref{section2}.
The multicritical point at $j=1$, where the magnetism is fully frustrated,
is our main concern.

In order to simulate the $(2+1)$-dimensional gonihedric model,
we utilize the numerical-diagonalization method.
This approach may have the following advantages.
First,
we implemented Novotny's method 
\cite{Novotny90}
to represent the Hamiltonian-matrix
elements;
this method is readily applicable to the quantum-mechanical system 
as well \cite{Nishiyama07}.
Owing to this method, we are able to treat an arbitrary number of
spins $N=8,12,\dots,28$ constituting the $d=2$ cluster;
note that conventionally, the number of spins is restricted 
within $N(=L^2)=9,16,25,\dots$.
Such an arbitrariness allows us to make a systematic finite-size scaling
analysis.
Second, the diagonalization method is free from the
slowing-down problem;
this problem becomes severe for such a frustrated magnetism,
deteriorating the efficiency of the Monte Carlo sampling.
Last, the constraint 
$L_\perp^z/L_\parallel \to 0$
[Eq. (\ref{constraint})] is always satisfied,
because 
the system size along the imaginary-time direction
is infinite $L_\parallel \to \infty$; note that the system size along the imaginary 
time corresponds to the inverse temperature
$L_\parallel=1/T\to\infty$.

In fairness, it has to be mentioned that our research
owes its basic idea to the following pioneering studies.
First, an equivalence between the
$(2+1)$-dimensional fully frustrated magnetism and the 
$(d,m)=(3,2)$ Lifshitz point was argued 
field-theoretically in 
Refs. \cite{Dutta97,Dutta98}.
Second, in Ref. \cite{Pal90}, the biaxial-next-nearest-neighbor Ising model
in $d=3$
was studied with the Monte Carlo method.
It was reported that 
the Lifshitz (multicritical) point 
collapses  at zero temperature.
On the contrary,
the gonihedric model has an extra tunable parameter $\kappa$.
Setting $\kappa \ge 2$,
we attain desirable multicriticality as depicted in Fig. \ref{figure1} (a).

The rest of this paper is organized
as follows.
In Sec. \ref{section2}, we explain the $(2+1)$-dimensional gonihedric model.
To elucidate the underlying physics,
we make an overview of the classical gonihedric model in $d=3$.
In Sec. \ref{section3}, we present the simulation results.
The simulation scheme is explained in the Appendix.
In Sec. \ref{section4}, we present a summary and discussions.

\section{\label{section2}
Quantum gonihedric model in $2+1$ dimensions:
A realization of the $(d,m)=(3,2)$ Lifshitz point}

In this section, we propose the $(2+1)$-dimensional gonihedric model
as a realization of the $(d,m)=(3,2)$ Lifshitz point.
To elucidate the underlying physics,
we make an overview of the original (classical) gonihedric model
in $d=3$.

\subsection{Quantum gonihedric model in $d=2$}

As mentioned in the Introduction,
we propose the $(2+1)$-dimensional gonihedric model
as a realization of
the
$(d,m)=(3,2)$ Lifshitz point.
To be specific, we consider the Hamiltonian
\begin{equation}
\label{Hamiltonian}
{\cal H}=
-J_1 \sum_{\langle ij \rangle} \sigma_i^z \sigma_j^z
-J_2 \sum_{\langle \langle ij \rangle \rangle} \sigma^z_i \sigma^z_j
-J_3 \sum_{[ijkl]} \sigma^z_i \sigma^z_j \sigma^z_k \sigma^z_l 
-\Gamma \sum_i \sigma^x_i   ,
\end{equation}
with the coupling constants 
$J_1=\kappa$,
$J_2=-\kappa/2$, and
$J_3=(1-\kappa)/2$.
Here, the operators $\{ \sigma^\alpha_i \}$ denote the Pauli matrices placed at
the square-lattice points $i$.
The summations 
$\sum_{\langle ij \rangle}$,
$\sum_{\langle \langle ij \rangle \rangle}$, and
$\sum_{ [ijkl] }$
run over all possible nearest-neighbor,
next-nearest-neighbor (plaquette diagonal),
and
plaquette-four-body spins, respectively.
The transverse magnetic field $\Gamma$ controls the amount of
quantum fluctuations.
At a certain $\Gamma_c$, a ground-state phase transition
may occur.

As mentioned above, the gonihedric model has finely tuned
coupling constants $\{ J_i \}$, which cancel out
the domain-wall surface tension.
Actually, the domain-wall energy of the gonihedric model
(apart from the off-diagonal term $-\Gamma \sum_i \sigma^x_i$)
admits a geometric representation $E=n_2+4\kappa n_4$ \cite{Savvidy94}.
Here, $n_2$ denotes the number of points where 
two domain walls meet at a right angle (domain-wall undulation),
and $n_4$ is the number of points where
four domain walls meet at a right angle (self-intersection point).
That is,
the parameter $\kappa$ controls the self-avoidance
of the domain walls
with the bending elasticity unchanged.
(Notably enough, the interfacial energy lacks 
the surface-tension term.
Accordingly, the domain-wall undulations
are promoted, giving rise
to a peculiar type of criticality.)
The gonihedric model has a 
tunable parameter $\kappa$
with the zero surface tension maintained.
This redundancy is an advantage over other 
frustrated magnetisms such as
the biaxial-next-nearest-neighbor Ising model.
We survey the regime $\kappa \ge 2$, where we observed
a clear indication of the Lifshitz-type criticality.

In this paper, we extend the above-mentioned parameterization space.
That is, introducing a new controllable parameter $j$,
we investigate the parameter space
\begin{equation}
\label{parameter_space}
J_1 = \kappa , \
J_2 = -\frac{\kappa  j }{2}  , \
J_3 = \frac{1-\kappa}{2}  .
\end{equation}
Note that at $j=1$, the parameter space, Eq. (\ref{parameter_space}), reduces to the above-mentioned one
(original gonihedric model).
Owing to the extension, the magnetic domain wall now acquires 
a finite domain-wall surface tension $\propto 1-j$.
In other worlds, in terms of this extended parameter space,
we identify the Lifshitz point as a multicritical point;
see the phase diagram in Fig. \ref{figure1} (a).
This viewpoint was proposed in Ref.  \cite{Cirillo97b},
where the authors investigate
the criticality of the
classical $d=3$ gonihedric model with the cluster-variation method.
In the next section,
we will overview the properties of the classical gonihedric model,
which may be relevant to the present study.

\subsection{Phase diagram of the classical gonihedric model:
A brief overview}

Let us make an overview of the past studies of
the (classical) gonihedric model.
The model was introduced by Savvidy and Wegner as a 
lattice-regularized version
of the string field theory \cite{Savvidy94,Ambartzumian92}.
However, recent developments dwell on the $d=3$ case,
aiming at a potential applicability to microemulsions.
The criticality should belong
to the Lifshitz point with the index $(d,m)=(3,3)$,
because the classical gonihedric model is isotropically frustrated.
The $(d,m)=(3,3)$
criticality may be realized by the ternary mixture \cite{Dawson88} 
of water, oil and surfactant
\cite{Huang81,Honorat84,Aschauer93,Seto96};
actually, a crossover from the $d=3$-Ising universality to
an exotic one was
reported in Refs. \cite{Schwahn99,Stepanek02}.

We present a schematic phase diagram of the
(classical) $d=3$ gonihedric model in Fig. \ref{figure1} (b)
\cite{Cirillo97b,Nishiyama04}.
The Hamiltonian of the classical $d=3$ gonihedric model is given by
\begin{equation}
\label{Hamiltonian_classical}
H=-\sum_{\langle ij \rangle} S_i S_j
  -j \sum_{\langle \langle ij \rangle\rangle} S_i S_j
  -\frac{1-\kappa}{4\kappa} \sum_{[ijkl]} S_i S_j S_k S_l  .
\end{equation}
(The Ising-spin variables $\{ S_i \}$ are placed at the $d=3$ lattice points.)
We notice that the phase diagram resembles that of the quantum 
gonihedric model;
the discrepancy as to $j \leftrightarrow -j$ is merely due to 
the difference of parameterization, and
the subspace $-j=1/4$ corresponds to the %  original
fully-frustrated gonihedric model.

A few remarks on the phase diagram follow:
First, the Lifshitz point at $-j=1/4$ is identified as an
end-point of the critical branch 
($-j < 1/4$) 
belonging to the $d=3$-Ising universality.
In fact,
the multicritical (crossover) scaling theory applies 
successfully \cite{Cirillo97b,Nishiyama04}
to clarifying the nature of the Lifshitz point.
(Direct numerical simulation at $-j=1/4$ appears to be
rather problematic
\cite{Hellmann93}.)
We will accept this cross-over viewpoint as
for the quantum gonihedric model.
Second, in Refs. \cite{Koutsoumbas02,Espriu97}, it was reported that
for small $\kappa < 0.5$,
the multicritical point becomes a discontinuous one,
accompanied with pronounced hysteresis.
In particular, at $\kappa=0$,
the model reduces to the so-called $p$-spin model
\cite{Shore91},
which is notorious for its slow relaxation to the thermal equilibrium
(metastability).
We found that a similar difficulty arises in the quantum gonihedric model.
Hence, we devote ourselves to the large-$\kappa$ regime such as $\kappa \ge 2$,
where we observed a clear indication of the Lifshitz-type criticality.
Last,
the phase boundary separating the lamellar and ferromagnetic phases
is (almost) vertical.
This feature ensures that the multicritical point is located at $-j=1/4$.
The quantum gonihedric model possesses this property as shown in the next section.
Actually, this is the most significant benefit of
the parameterization scheme, Eq. 
(\ref{parameter_space}).

\section{\label{section3}Numerical results}

In this section, we present the numerical results.
Our aim is to estimate the critical indices 
$(\nu_\perp,\nu_\parallel)$ and $\phi$.
As mentioned in the Introduction,
we utilize Novotny's method to diagonalize the Hamiltonian 
(\ref{Hamiltonian}) numerically.
We explain the technical details in the Appendix.
By means of this method,
we simulated finite clusters with $N \le 28$ spins.
The linear dimension of the cluster $L$ is given by the formula
\begin{equation}
L=\sqrt{N}   ,
\end{equation}
because the $N$ spins constitute a $d=2$ cluster.

\subsection{\label{section3_1}Finite-size scaling of the critical branch: $d=3$-Ising universality}

In this section, we survey the critical branch $j<1$; see Fig. \ref{figure1} (a).
We show that the criticality belongs to the ordinary
$d=3$-Ising universality class.
This finding provides a foundation for the subsequent analyses
with the crossover-scaling theory.

\begin{figure}
\includegraphics[width=100mm]{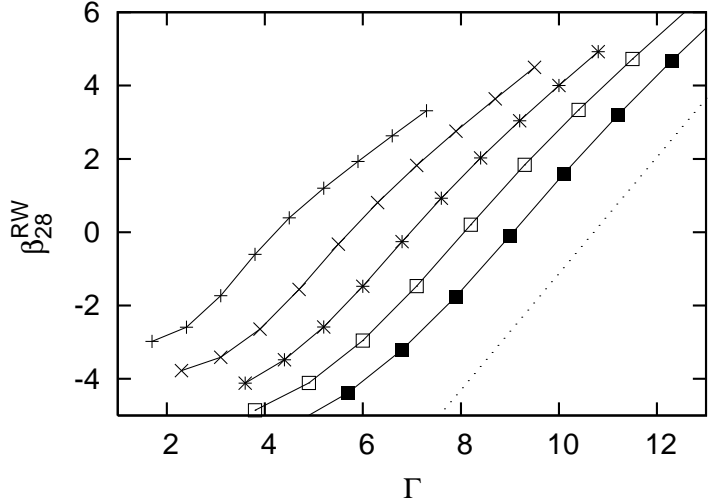}%
\caption{
\label{figure2}
The beta function $\beta^{RW}_{28}(\Gamma)$ (\ref{beta_function})
is plotted for $\kappa=2$ and various $j$.      %%% 4
The symbols,
$+$, $\times$, $*$, $\Box$, and $\blacksquare$,
denote the data for
$j=0.4$,
$0$, $-0.4$, $-0.8$, and $-1.2$, respectively.   % .2 0 -.2 -.4
For a comparison,
we presented a slope
(dotted line) corresponding to the
$d=3$-Ising universality class ($\nu=0.6294$ \cite{Deng03}).
We see that the criticality is maintained to be
that of the $d=3$-Ising universality class for a wide
range of $j$; see text for details.
}
\end{figure}

In Fig. \ref{figure2},
we plot the Roomany-Wyld approximate beta function
\cite{Roomany80}
\begin{equation}
\label{beta_function}
\beta^{RW}_N(\Gamma)=
\frac{1+\ln(\Delta E_N(\Gamma)/\Delta E_{N-4}(\Gamma))/
                        \ln (\sqrt{N}/\sqrt{N-4})}
     {
\sqrt{\partial_\Gamma \Delta E_N(\Gamma) \partial_\Gamma \Delta E_{N-4}(\Gamma)/
     \Delta E_N(\Gamma) / \Delta E_{N-4}(\Gamma)}}
          ,
\end{equation}
with the excitation energy gap $\Delta E_N(\Gamma)$ for the system size $N$.
Here, we fixed the self-avoidance parameter $\kappa=2$, and varied
%%%  k=4  the frustration as $j=-0.4$, $-0.2$, $0$, $0.2$, and $0.4$.
the frustration as $j=-1.2$, $-0.8$, $-0.4$, $0$, and $0.4$.
The zero point of the beta function
indicates the location of the critical point $\Gamma_c(j)$.
Basically,
the critical branch depicted in Fig. \ref{figure1} (a) 
follows from this analysis;
afterward,
we determine the critical point $\Gamma_c$ more precisely.
%(The first-order branch in $j>1$ is rather out of our concern.)

The slope of the beta function at $\Gamma=\Gamma_c$
yields an estimate for the inverse of the correlation-length
critical exponent, $1/\nu$.
In Fig. \ref{figure2}, as a reference,
we presented a slope (dotted line) 
$1/\nu=1.5868$ \cite{Deng03}
corresponding to the
$d=3$-Ising universality class.
We see that the criticality is maintained in
the $d=3$-Ising universality class 
for a wide range of $j$.
Actually,
we obtained   %   at $\Gamma=\Gamma_c(j)$
%$1/\nu= 1.63$, $1.61$,  $1.61$,  $1.63$, and $1.84$
%for $j=-0.4$, $-0.2$, $0$, $0.2$, and $0.4$,
$1/\nu= 1.55$, $1.54$,  $1.51$,  $1.46$, and $1.39$
for $j=-1.2$, $-0.8$, $-0.4$, $0$, and $0.4$,
respectively.
These results demonstrate that the critical branch
belongs to the $d=3$-Ising universality class.
%#From this observation,
%#we estimate the exponent along the critical branch as $1/\nu$
%#in good agreement with the $d=3$-Ising universality class.
It would be noteworthy that
the shape of the beta function becomes distorted as $j \to 1$.
That is,
the regime exhibiting the slope $1/\nu$ shrinks gradually
as $j \to 1$, indicating that a new type of criticality
emerges at the multicritical point $j=1$.
Actually, we consider
this crossover behavior rather in detail
in the following sections.

\begin{figure}
\includegraphics[width=100mm]{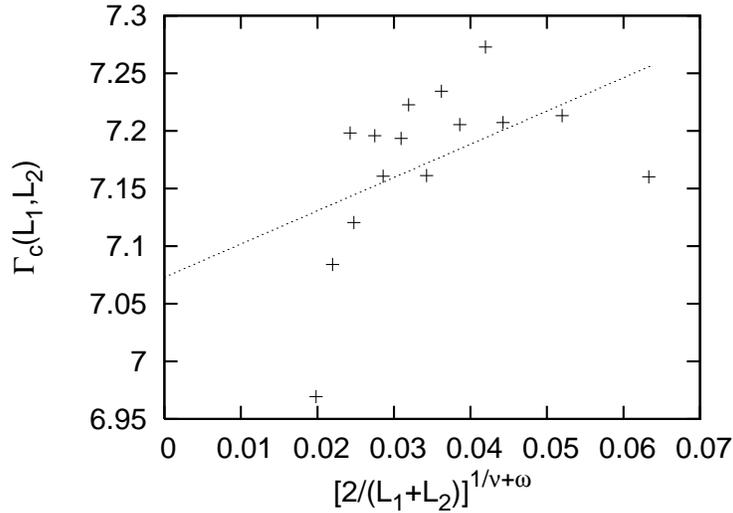}%
\caption{
\label{figure3}
The approximate critical point 
$\Gamma_c(L_1,L_2)$ is plotted for 
$[2/(L_1+L_2)]^{1/\nu+\omega}$
with $8 \le N_1 < N_2 \le 28$, $\kappa=2$,       %    4$,
and
$j=-0.4$;   %%%    0$;
the corrections-to-scaling exponent $\omega=0.821$
and the exponent $1/\nu=1.5868$ are taken from
Ref. \cite{Deng03}.
The least-squares fit to these data yields
$\Gamma_c=7.073(55)$      %    10.41(21)$
in the thermodynamic limit.
}
\end{figure}

In Fig. \ref{figure3},
we present the approximate critical point
$\Gamma_c(L_1,L_2)$
for $[2/(L_1+L_2)]^{1/\nu+\omega}$ \cite{Binder81} with 
$\kappa=2$, $j=-0.4$, and     %%   4 0
$8 \le N_1<N_2 \le 28$ ($L_{1,2}=\sqrt{N_{1,2}}$);
here, we used the corrections-to-scaling exponent 
$\omega=0.821$
and the exponent
$\nu=1.5868^{-1}$ reported in Ref. \cite{Deng03}.
The approximate critical point
$\Gamma_c(L_1,L_2)$
 is determined by the zero point of
the beta function.
That is, it satisfies the equation 
\begin{equation}
\sqrt{N_1} \Delta E_{N_1}(\Gamma_c(L_1,L_2))=
\sqrt{N_2} \Delta E_{N_2}(\Gamma_c(L_1,L_2))   .
\end{equation}
From the least-squares fit to the data in Fig. \ref{figure3}, 
we obtained the critical point 
%%$\Gamma_c=10.41(21)$
$\Gamma_c=7.073(55)$
in the thermodynamic limit $L \to \infty$.
We make use of $\Gamma_c$ in the following scaling analyses.

\subsection{\label{section3_2}End-point singularity of the critical amplitude}

The above analysis indicates that the multicriticality at $j=1$ 
is merely an end-point singularity of the ordinary $d=3$-Ising critical branch.
That is, the crossover-scaling theory should apply to clarifying the
nature of the multicritical point.

In this section,
we consider the singularity of the critical amplitude of $\Delta E$
beside the multicritical point.
The amplitude $G^{\pm}$ is defined by the relation
\begin{equation}
\Delta E \approx G^{\pm}(j) |\Gamma-\Gamma_c(j)|^{\nu}   .
\end{equation}
The amplitude exhibits the singularity
\begin{equation}
\label{amplitude_singularity}
G^{\pm} (\Delta)  \sim \Delta^{(\nu_\parallel-\nu)/\phi}   ,
\end{equation}
with the crossover exponent $\phi$.
(As mentioned in the Introduction,
the exponent $\nu_{\parallel}$ denotes the critical index 
along the imaginary-time direction.)
The variable $\Delta$ stands for the distance from the
multicritical point 
\begin{equation}
\Delta=1-j  .
\end{equation}
Here, we postulated that the multicritical point locates at $j=1$,
and we justify this claim in Sec. \ref{section3_4}.
The above formula is a straightforward consequence of the crossover-scaling hypothesis
\begin{equation}
\label{crossover_scaling}
\Delta E \approx |\Gamma-\Gamma_c|^{\nu_\parallel} f(\Delta/|\Gamma-\Gamma_c|^\phi) .
\end{equation}
Actually,
this relation provides a definition of the crossover exponent $\phi$.

\begin{figure}
\includegraphics[width=100mm]{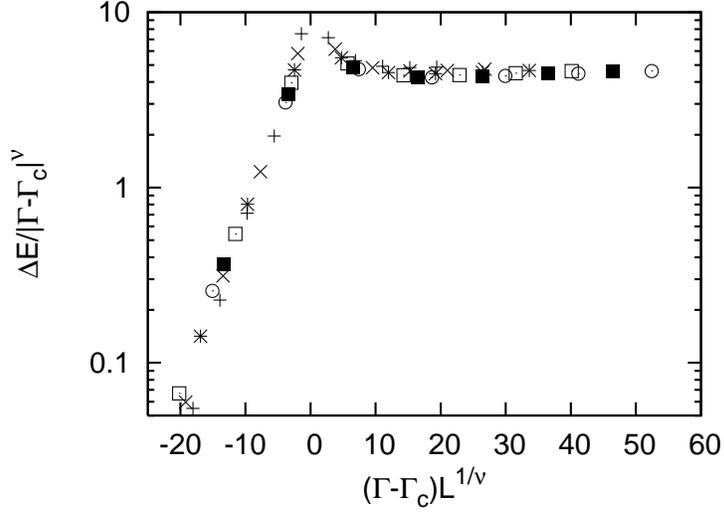}%
\caption{
\label{figure4}
The scaling plot for the energy gap,
$(\Gamma-\Gamma_c) L^{1/\nu}$-
$\Delta E  /  |\Gamma-\Gamma_c|^{\nu}$,
is shown.  % for $\kappa=2$ and $j=-0.4$.     % 4   0
The parameters are the same as those of Fig. \ref{figure3}.
We postulated the $d=3$-Ising universality class $\nu=0.6302$ \cite{Deng03}. 
The symbols,
$+$, $\times$, $*$, $\Box$, $\blacksquare$, and $\circ$,
denote the system sizes of $N=8$, 12, 16, 20, 24, and 28, respectively.
We confirm that the transition belongs to the $d=3$-Ising universality class.
Furthermore,
from the plateau in the high-$\Gamma$ side, we obtain an estimate for the
critical amplitude $G^+=4.28(8)$; see text for details.
                    %%%%%% 4.52(16)
}
\end{figure}

To begin with, we determine the critical amplitude $G^+$.
In Fig. \ref{figure4},
we plot the scaled energy gap 
$(\Gamma-\Gamma_c)L^{1/\nu}$-$\Delta E / |\Gamma-\Gamma_c|^{\nu}$
for $\kappa=2$, $j=-0.4$, and $N=8,12,\dots,28$.   %%    4 0
The critical point $\Gamma_c=7.073$ is determined in the above section,    % 10.41
and likewise, we postulated the $d=3$-Ising universality class $1/\nu=1.5868$ \cite{Deng03}.
The data collapse into a scaling-function curve.
We again confirm that the phase transition belongs to the
$d=3$-Ising universality class.
From the limiting value of the
high-$\Gamma$ side of the scaling function,
we estimate the critical amplitude
as $G^+ =  4.28(8)$;               %%  4.52(16)$;
here, we read off the value around the scaling regime 
$(\Gamma-\Gamma_c)L^{1/\nu}=15$,
and accepted the data scatter among $N=20$, $24$ and $28$ as
an error indicator.

\begin{figure}
\includegraphics[width=100mm]{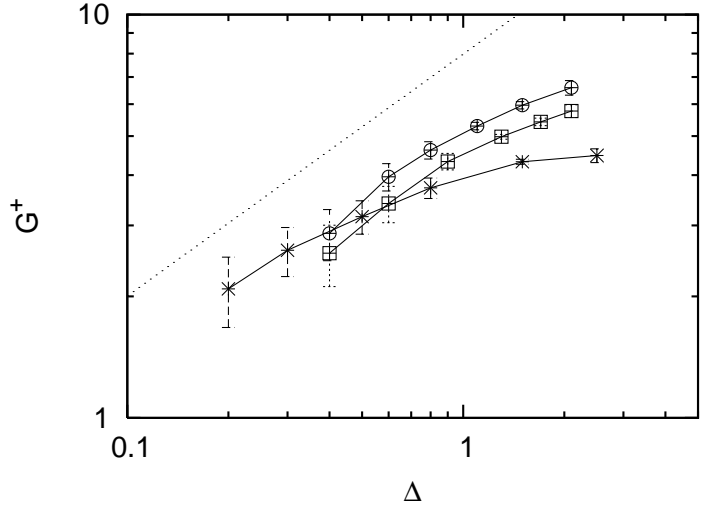}%
\caption{
\label{figure5}
The
correlation-length critical amplitude $G^+$ 
(\ref{amplitude_singularity})
is plotted for various $\Delta(=1-j)$ and $\kappa=2$, 4, and 6.
The symbols 
$\times$, $\Box$, and $\circ$
denote the data for $\kappa=2$, 4, and 6, respectively.
As a reference, we presented a slope (dotted line)
of $G^+ \propto \Delta^{0.6}$.
The data indicate a power-law singularity, Eq. 
(\ref{amplitude_singularity}),
with the exponent 
$(\nu_\parallel-\nu)/\phi \approx 0.6$;
see text for details.
}
\end{figure}

Similarly, we determined $G^+$ for various values of 
$j$ and
$\kappa=2$, 4, and 6.
In Fig. \ref{figure5},
we plotted the amplitude $G^+$ for $\Delta(=1-j)$
with the logarithmic scale.
[In the cases of $\kappa=2$, $4$, and $6$,
we read off $G^+$ from the scaling plot at the
scaling regime $(\Gamma-\Gamma_c)L^{1/\nu}=15$, 40, and 60, respectively.
In the case of $\kappa=2$, 
we omitted the data of $N=16$ for its
rather insystematic behavior particularly for small $\Delta$.]
In the plot, we also presented a slope (dotted line) of $G^+ \propto \Delta^{0.6}$.
We observe a signature of the
power-law singularity with the exponent 
$(\nu_\parallel -\nu)/\phi   \approx 0.6$
as $\Delta \to 0$.
Hence, we confirm that the crossover behavior (\ref{amplitude_singularity})
is realized in the vicinity of the multicritical point.
In fact, 
from 
$\nu=0.63020(12)$ \cite{Deng03}
and the present results,
Eqs. 
(\ref{exponent_nu_parallel}) and 
(\ref{exponent_phi}), obtained in Sec. \ref{section3_3},
we arrive at the slope
\begin{equation}
\frac{\nu_\parallel-\nu}{\phi}=0.59(42)  ,
\end{equation}
fairly consistent with the above observation.
With use of $G^+$ calculated 
in this section,
we crosscheck the validity of the
critical indices
obtained in the following section.

\subsection{\label{section3_3}
Finite-size-scaling analysis of $(\nu_\perp,\nu_\parallel)$ and $\phi$}

In this section, we make an analysis of
each critical exponent
with use of the crossover scaling, Eq. (\ref{crossover_scaling}).

\begin{figure}
\includegraphics[width=100mm]{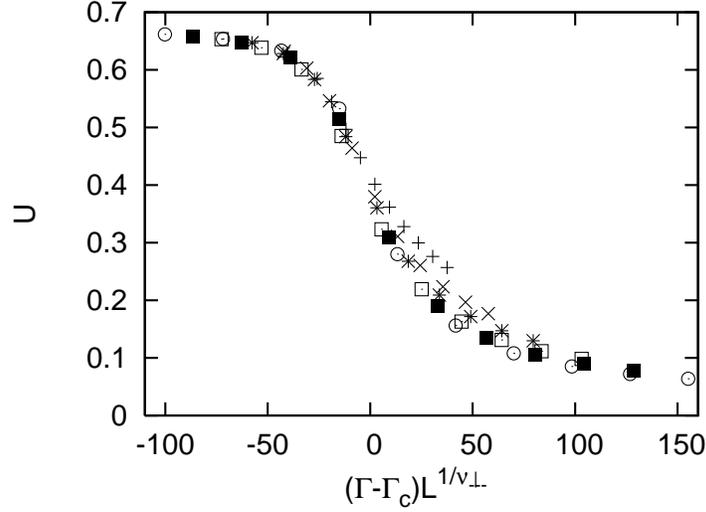}%
\caption{
\label{figure6}
The crossover-scaling plot 
(\ref{scaling_Binder}),
$(\Gamma-\Gamma_c)L^{1/\nu_\perp}$-$U$,
for $\kappa=2$ and $\Delta L^{\phi/\nu_\perp}=8$ is shown.     %% 4 10
Here we set $\nu_\perp=0.45$ and $\phi=0.7$, for which we found the best data collapse.
The symbols,
$+$, $\times$, $*$, $\Box$, $\blacksquare$, and $\circ$,
denote the system sizes of $N=8$, 12, 16, 20, 24, and 28, respectively.
}
\end{figure}

First, we consider the Binder parameter
\begin{equation}
U=1-
  \frac{\langle M^4 \rangle}{3 \langle M^2 \rangle^2}   ,
\end{equation}
with the magnetization 
$M=\sum_{i=1}^N \sigma^z_i$.
  [Note that the simulation was not done right at the Lifshitz
  point; we calculated the data in the vicinity of the Lifshitz
  point (crossover scaling). Hence, the ferromagnetic order 
  parameter $M$ is still of use in the data analysis.]
The symbol
$\langle \dots \rangle$
denotes the expectation value at the ground state.
According to
the crossover-scaling theory,
the Binder parameter obeys the formula
\begin{equation}
\label{scaling_Binder}
U=\tilde{U}(  (\Gamma-\Gamma_c) L^{1/\nu_\perp},\Delta L^{\phi/\nu_\perp})
.
\end{equation}
(Here, we made use of the fact that
the Binder parameter is scale-invariant at the critical point.)
As noted in the Introduction, the index with the subscript
$\perp$ denotes the critical exponent within the real space.
In Fig. \ref{figure6},
we present the crossover-scaling plot, 
$(\Gamma-\Gamma_c)L^{1/\nu_\perp}$-$U$, with $\kappa=2$ and %  the fixed     %%%  4
$\Delta L^{\phi/\nu_\perp}=8$.      %  10
Here, we set the scaling parameters $\nu_\perp=0.45$ and $\phi=0.7$,
where we found the best data collapse.
Surveying $\kappa=4$ and $6$ as well,
we arrive at the estimates 
\begin{equation}
\label{exponent_nu_perp}
\nu_\perp  = 0.45(10) ,
\end{equation}
and
\begin{equation}
\label{exponent_phi}
\phi = 0.7(2)   .
\end{equation}

\begin{figure}
\includegraphics[width=100mm]{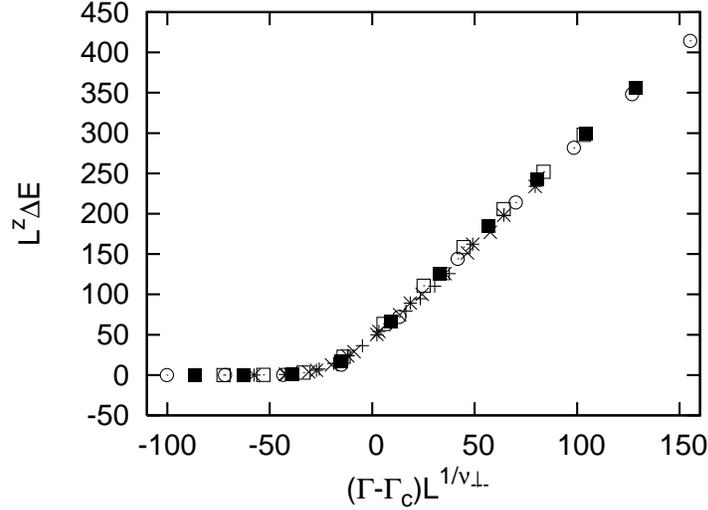}%
\caption{
\label{figure7}
The crossover-scaling plot 
(\ref{scaling_gap}),
$(\Gamma-\Gamma_c)L^{1/\nu_\perp}$-$L^z \Delta E $, is shown.
Here, we set $z=2.3$, and the other scaling parameters are the same
as those of Fig. \ref{figure6}.
The symbols,
$+$, $\times$, $*$, $\Box$, $\blacksquare$, and $\circ$,
denote the system sizes of $N=8$, 12, 16, 20, 24, and 28, respectively.
}
\end{figure}

Second, we consider the energy gap $\Delta E$.
The energy gap obeys the crossover-scaling relation
\begin{equation}
\label{scaling_gap}
\Delta E=
L^{-z}
%%|\Gamma-\Gamma_c|^{\nu_\parallel} 
%%   f(\Delta/|\Gamma-\Gamma_c|^\phi)
g(  (\Gamma-\Gamma_c) L^{1/\nu_\perp},\Delta L^{\phi/\nu_\perp})  ,
\end{equation}
with the dynamical critical exponent $z$.
In Fig. \ref{figure7},
we present the crossover-scaling plot, 
$(\Gamma-\Gamma_c)L^{1/\nu_\perp}$-$L^z \Delta E$, with $\kappa=2$ and % the fixed   %% 4
$\Delta L^{\phi/\nu_\perp}=8$.    %%%% 10
Here, we set $z=2.3$, and the other 
scaling parameters are the same as those of Fig. \ref{figure6}.
Surveying $\kappa=4$ and $6$ as well,
we estimate the critical index as
\begin{equation}
z=2.3(3)    .
\end{equation}
Through $z=\nu_\parallel /\nu_\perp$,
the above results lead to
\begin{equation}
\label{exponent_nu_parallel}
\nu_\parallel=1.04(27)    .
\end{equation}

 Let us address a remark.
As mentioned in the above section,
the indices, Eqs. 
(\ref{exponent_phi}) and 
(\ref{exponent_nu_parallel}), are consistent with the 
end-point singularity
of $G^+$,
indicating the self-consistency of the present analyses.

\subsection{\label{section3_4}Phase transition between the ferromagnetic and lamellar phases}

The above analysis
stems from the proposition that the multicritical point locates at $j=1$;
in other words, the phase boundary separating the ferromagnetic and lamellar phases
is (almost) vertical.
In this section, we justify this proposition.
(Actually, this feature was confirmed in the case of 
the classical $d=3$ gonihedric model \cite{Cirillo97b}.)

\begin{figure}
\includegraphics[width=100mm]{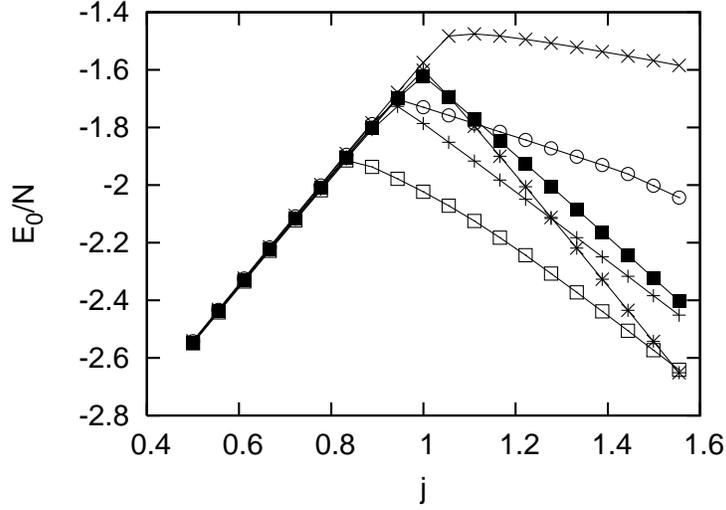}%
\caption{
\label{figure8}
The ground-state energy per unit cell $E_0/N$
is plotted for $\kappa=2$ and $\Gamma=0.6$.
The symbols,
$+$, $\times$, $*$, $\Box$, $\blacksquare$, and $\circ$,
denote the system sizes of $N=8$, 12, 16, 20, 24, and 28, respectively.
There occurs a transition separating the ferromagnetic
and lamellar phases.
The transition point seems to 
converge into the regime
$0.9 \le j_c \le 1$ as $N \to \infty$;
see text for details,
%Low lying energy spectrum is shown for $\kappa=4$ and
%$\Gamma=$; see text for details.
%We see an indication of the first-order phase transition
%separating the ferromagnetic and lamellar phases;
%see Fig. \ref{figure1} (a).
%This result indicates that the phase boundary 
%locates at $j \approx 1$.
}
\end{figure}

In Fig. \ref{figure8}, we plot the ground-state energy per unit cell
$E_0/N$ with the system sizes $N=8,12,\dots,28$ for
$\kappa=2$ and $\Gamma=0.6$; namely, we surveyed the regime slightly below the
multicritical point.
We observe a distinct signature of the first-order phase transition around $j \approx 1$,
where the slope of $E_0/N$ changes rather abruptly (level crossing).
The transition point seems to converge into
the regime
$0.9  \le j_c \le 1$ as $N \to \infty$.
Noticeably enough, the transition point is close to $j=1$.

We argue this behavior more in detail:
First,
the data $E_0 / N$ in $j< j_c$ (ferromagnetic phase)
appear to reach the thermodynamic limit, 
whereas 
in $j>j_c$ (lamellar phase),
the plots are still scattered insystematically.
Possibly,
the incommensurability of the lamellar-type structure (periodicity of the domain walls)
causes such an irregularity.
Surveying the cases of $\kappa=2$, $4$, and $6$,
we found that the data of $N=8$, 16, and 24 are rather robust 
against this incommensurability effect.
Hence, we conclude
that the transition point locates within $0.9 \le j_c \le 1$.
%note that the parameter $\kappa$ controles the self-avoidance of the domain walls,
%and it enhances the incommesurability effect.
%%Second,
Last,
we found that such a slight deviation of $j_c$ from $j=1$ is negligible in the
 sense that 
the influence is less than the error margins.
In other worlds, the parameterization, Eq. (\ref{parameter_space}),
is sensible to explore the multicriticality 
in terms of the cross-over scaling;
this point was noted in the case of the
classical gonihedric model 
\cite{Cirillo97b}.
%Last,
%the incommensurability may be the source of the difficulty
%of the direct simulation at $j=1$ \cite{Hellmann93}.
%Rather, the crossover-scaling analysis, albeit elaborative,
%is suitable to explore the Lifshitz criticality.

\section{\label{section4}Summary and discussions}

We investigated
the criticality of the $(2+1)$-dimensional gonihedric model,
Eq.
(\ref{Hamiltonian}), with the extended parameter space, Eq.
(\ref{parameter_space}).
This extended parameter space allows us to survey the criticality
in terms of the crossover-scaling theory; see Fig. \ref{figure1} (a).
We employed Novotny's method to diagonalize the Hamiltonian.
%%%(\ref{Hamiltonian}).
With use of this method, we treated
an arbitrary (integral) number of spins $N=8,12,\dots,28$.
Because the quantum-mechanical fluctuation along the imaginary time direction
is ferromagnetic, the criticality of the quantum gonihedric model
should be an anisotropic one accompanied with the dynamical critical exponent 
$z(=\nu_\parallel/\nu_\perp) \ne 1$.
Our estimates for the critical indices are 
$(\nu_\perp,\nu_\parallel)=(0.45(10),1.04(27))$ and $\phi=0.7(2)$.
We also confirmed that the estimates are consistent with
the end-point singularity of the
critical amplitude $G^+$; see Fig. \ref{figure5}

As mentioned in the Introduction,
Diehl and Shpot made an analysis of the Lifshitz point with the
$\epsilon$-expansion method up to O$(\epsilon^2)$.
Their conclusions for $(d,m)=(3,2)$ are
\begin{equation}
(\nu_\perp,\nu_\parallel)=(0.387,0.795)   \ and \ \phi=0.686  .
\end{equation}
They also provided the convergence-accelerated results with 
the $[1/1]$ Pad\'e method;
\begin{equation}
(\nu_\perp,\nu_\parallel)=(0.482,1.230)  \ and \ \phi=0.688 .
\end{equation}
Our simulation data support their claim.
%Particularly, our data agree with the Pad\'e results;
%however, as to $\nu_\parallel$ (equivalently $z$),
%our simulation result suggests a slightly moderate value.

Lastly, let us make a few comments on the advantages of the
diagonalization approach.
First, the numerical diagonalization is free from the
slowing-problem problem, which deteriorates the efficiency of
the Monte Carlo sampling for the frustrated magnetism.
Second,
we do not have to worry about the constraint 
(\ref{constraint}).
The constraint is always satisfied,
because the system size along the imaginary-time direction is infinite.
However,
the diagonalization method suffers from the
severe limitation as to the available system sizes.
In this paper, we surmount this difficulty
with the aide of Novotny's method,
which allows us to treat a variety of system sizes
$N=8,12,\dots,28$ sufficient to manage systematic
finite-size scaling.

\begin{acknowledgments}
This work is supported by a Grant-in-Aid 
(No. 18740234) from Monbu-Kagakusho, Japan.
\end{acknowledgments}

\appendix

\section{
Construction of the Hamiltonian-matrix elements: Quantum Novotny's method}

In this Appendix, we explain the simulation scheme.
As mentioned in the Introduction,
we applied the Novotny method \cite{Novotny90} to diagonalizing the Hamiltonian
(\ref{Hamiltonian}).
Novotny's method allows us to construct the Hamiltonian-matrix
elements systematically for the cluster with
an arbitrary (integral) number of spins 
$N=8,12,\dots,28$;
note that conventionally, the number of spins is restricted within 
$N(=L^2)=9,16,25,\dots$.
Originally,
Novotny's method was formulated for the classical 
Ising model (transfer-matrix formalism) \cite{Novotny90}.
In Ref. \cite{Nishiyama07},
it was extended to adopt the quantum-mechanical interaction (Hamiltonian formalism).
Here, we follow the notation of Ref. \cite{Nishiyama07}, 
and make a slight extension 
to incorporate the plaquette-type interactions;
see Eq. (\ref{plaquette_interaction}).

\begin{figure}
\includegraphics[width=100mm]{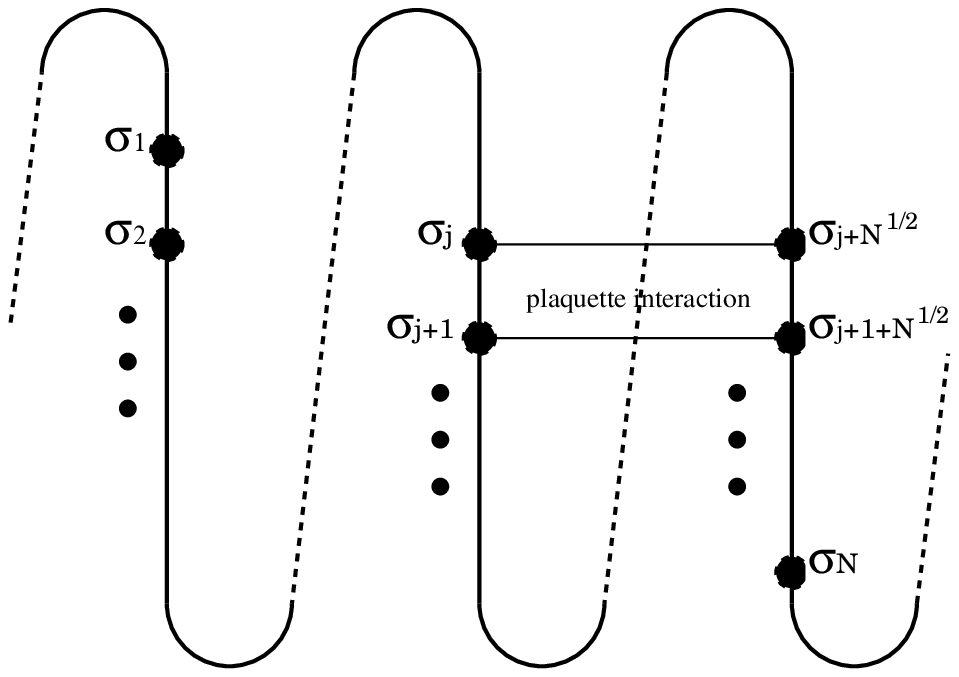}%
\caption{
\label{figure9}
Construction of the spin cluster for the
gonihedric model (\ref{Hamiltonian}).
As indicated above,
the spins constitute a $d=1$-dimensional alignment 
$\{ \sigma_i\}$ ($i=1,2,\dots,N$),
and the dimensionality is lifted to $d=2$ by introducing
the bridges
(long-range interactions) over the ($N^{1/2}$)th-neighbor pairs.
}
\end{figure}

Before we commence a detailed discussion,
we explain the basic idea of Novotny's method.
In Fig. \ref{figure9}, 
we present a schematic drawing of a finite-size cluster for
the $d=2$ gonihedric model, Eq. (\ref{Hamiltonian}).
As seen in the figure,
the spins $\{  \sigma_i   \}$ ($i=1, 2, \dots ,N$) constitute a 
$d=1$-dimensional (zig-zag)
structure.
This feature is essential for
us to construct the cluster
with an arbitrary (integral) number of spins $N$.
The dimensionality is lifted to $d=2$ by the long-range interactions
over the $\sqrt{N}$th-neighbor distances;
owing to the long-range interaction, the $N$ spins constitute a
$\sqrt{N}\times \sqrt{N}$
rectangular network effectively.
(The significant point is that the number $\sqrt{N}$ is not necessarily
an integral nor rational number.)

Let us formulate the above idea explicitly.
To begin with, we set up the Hilbert-space bases 
$\{| \sigma_1,\sigma_2,\dots,\sigma_N \rangle \}$
($\sigma_i = \pm 1$) for the quantum spins 
$\{ \sigma_i^\alpha \}$ ($i=1,2,\dots,N$).
The bases diagonalize the operator $\sigma^z_i$;
namely, the relation
\begin{equation}
\sigma_j^z | \{ \sigma_i \} \rangle  =
\sigma_j   | \{ \sigma_i \} \rangle  ,
\end{equation}
holds.

We decompose the Hamiltonian into two components
\begin{equation}
{\cal H}={\cal H}^{(D)}(\{ J_i \})
     +{\cal H}^{(O)}(\Gamma) .
\end{equation}
The component ${\cal H}^{(D)}(\{J_i\})$ describes the exchange interactions,
depending on the coupling constants $\{J_i\}$.
On the other hand, the component ${\cal H}^{(O)}(\Gamma)$ originates from
the single-spin term, which depends on the transverse magnetic field
$\Gamma$.
The former component is a diagonal matrix,
whereas the latter is off-diagonal.

First, we consider the diagonal component ${\cal H}^{(D)}$. 
We propose the following formula \cite{Nishiyama07}
\begin{equation}
\label{symmetrization}
{\cal H}^{(D)}=  \frac{1}{2}(H(\sqrt{N})+H(-\sqrt{N}))
   .
\end{equation}
Here, the component 
$H(v)$ is a diagonal matrix, which 
describes the $v$th-neighbor interaction among the $N$-spin alignment.
The diagonal elements are given by
\begin{equation}
\label{TP_decomposition}
H_{ \{\sigma_i\},\{\sigma_i\} }(v)
=\langle \{\sigma_i\} | H(v) | \{\sigma_i\} \rangle
=\langle \{\sigma_i\} | TP^v | \{\sigma_i\} \rangle
   .
\end{equation}
Here, the matrix $T$ denotes
the plaquette-type interaction
between the arrays
$\{ \sigma_i \}$ and $\{ \tau_i \}$;
\begin{equation}
\label{plaquette_interaction}
\langle \{\sigma_i\} |T| \{\tau_i\} \rangle=
\sum_{k=1}^{N}
\left(
-\frac{J_1}{2}(\sigma_k \sigma_{k+1}+\tau_k \tau_{k+1}
 +\sigma_{k}\tau_{k}+\sigma_{k+1}\tau_{k+1})
-J_2 (\sigma_k\tau_{k+1}+\sigma_{k+1}\tau_k)
-J_3 \sigma_k\sigma_{k+1}\tau_k\tau_{k+1}
\right)   .
\end{equation}
The operator $P$ denotes
the translational operator, which satisfies
$P|\{\sigma_i\}\rangle=|\{\sigma_{i+1}\}\rangle$;
here,
we imposed the periodic-boundary condition.
Note that the operator insertion of $P^v$ 
in Eq. (\ref{TP_decomposition}) introduces the long-range interaction
over the $v$th-neighbor pairs.
The denominator $2$ in Eqs.
(\ref{symmetrization}) and 
(\ref{plaquette_interaction}) compensates the duplicated sum.

Lastly, we consider the off-diagonal component ${\cal H}^{(O)}$.
The matrix element is given by
\begin{equation}
{\cal H}^{(O)}_{ \{\sigma_i\},\{\tau_i\} }=
\langle \{\sigma_i\} |
{\cal H}^{(O)}
| \{\tau_i\} \rangle
   .
\end{equation}
The expression is quite standard, because the component ${\cal H}^{(O)}$ simply concerns the
individual spins, and has nothing to do with the connectivity among them.

The above formulas complete our basis to simulate the Hamiltonian 
(\ref{Hamiltonian})
numerically.
The results are shown in Sec. \ref{section3}.

% Create the reference section using BibTeX:

\begin{thebibliography}{99}

%\input{bib.tex}
\bibitem{Diehl00}
H. W. Diehl and M. Shpot,
Phys. Rev. B {\bf 62}, 12338 (2000).
\bibitem{Shpot01}
M. Shpot and H. W. Diehl, Nucl. Phys. B {\bf 612}, 340 (2001).


\bibitem{Diehl03}
H. W. Diehl and M. Shpot,
Phys. Rev. B {\bf 68}, 066401 (2003).
\bibitem{Leite03a}
M. M. Leite, Phys. Rev. B {\bf 67}, 104415 (2003).
\bibitem{Leite03b}
M. M. Leite, Phys. Rev. B {\bf 68}, 066402 (2003).

%MA Shpot Yu M Pis'mak HW Diehl
%J Phys.: Condens. Matter 17 (2005) S1947
\bibitem{Hornreich75}
R. M. Hornreich, M. Luban, and S. Shtrikman,
Phys. Rev. Lett. {\bf 35}, 1678 (1975).
\bibitem{Mergulhao98}
C. Mergulh\~ao, Jr and C. E. I.  Carneiro,
Phys. Rev. B {\bf 58}, 6047 (1998).
\bibitem{Mergulhao99}
C. Mergulh\~ao, Jr and C. E. I.  Carneiro,
Phys. Rev. B {\bf 59}, 13954 (1999).




%
\bibitem{Selke78}
W. Selke, Z. Phys. B {\bf 29}, 133 (1978).
\bibitem{Kaski85}
K. Kaski and W. Selke, Phys. Rev. B {\bf 31}, 3128 (1985).
\bibitem{Pleimling01}
M. Pleimling and M. Henkel, Phys. Rev. Lett. {\bf 87}, 125702 (2001).
% series
\bibitem{Oitmaa85}
J. Oitmaa, J. Phys. A: Math. Gen. {\bf 18}, 365 (1985).
\bibitem{Mo91}
Z. Mo and M. Ferer, Phys. Rev. B {\bf 43}, 10890 (1991).



\bibitem{Savvidy94}G.K. Savvidy and F.J. Wegner, Nucl. Phys. B {\bf 413}, 605 (1994).
%\bibitem{Savvidy94b}G.K. Savvidy and K.G. Savvidy, Phys. Lett. B {\bf 324}, 72 (1994).
%\bibitem{Savvidy94c}G.K. Savvidy and K.G. Savvidy, Phys. Lett. B {\bf 337}, 333 (1994).
%\bibitem{Savvidy96}G.K. Savvidy, K.G. Savvidy, and P.G. Savvidy, Phys. Lett. A {\bf 221}, 233 (1996).
%%GK Savvidy KG Savvidy FJ Wegner
%%NP B 443 565 (1995)
%G Koutsoumbas GK Savvidy KG Savvidy  %MC
%PLB 410 241 (1997)
%
%\bibitem{Cappi93}A. Cappi, P. Colangelo, G. Gonnella, and A. Maritan, 
%Nucl. Phys. B {\bf 370}, 659 (1993).

\bibitem{Ambartzumian92}R.V. Ambartzumian, G.S. Sukiasian, G.K. Savvidy, and K.G. Savvidy,
Phys. Lett. B {\bf 275}, 99 (1992).


\bibitem{Cirillo97b}E.N.M. Cirillo, G. Gonnella, and A Pelizzola, Phys. Rev. E {\bf 55}, R17 (1997).


\bibitem{Koutsoumbas02}G. Koutsoumbas and G.K. Savvidy, Mod. Phys. Lett. A {\bf 17}, (2002) 751.
% 1st
\bibitem{Espriu97}D. Espriu, M. Baig, D.A. Johnston, and Ranasinghe P.K.C. Malmini,
J. Phys. A: Math. Gen. {\bf 30}, 405 (1997).




%
% Novotny
\bibitem{Novotny90}M.A. Novotny, J. Appl. Phys. {\bf 67}, 5448 (1990).
%\bibitem{Novotny92}M.A. Novotny, Phys. Rev. B {\bf 46}, 2939 (1992).
%\bibitem{Novotny93}M.A. Novotny, Phys. Rev. Lett. {\bf 70}, 109 (1993).
%\bibitem{Novotny91}M.A. Novotny, 
%{\it Computer Simulation Studies in Condensed Matter Physics III},
%edited by D.P. Landau, K.K. Mon, and H.-B. Sch\"uttler 
%(Springer-Verlag, Berlin, 1991).

%%\bibitem{Nishiyama05}Y. Nishiyama, Phys. Rev. E {\bf 71}, 046112 (2005).
%%\bibitem{Nishiyama06}Y. Nishiyama, Phys. Rev. E {\bf 73}, 016114 (2006).


\bibitem{Nishiyama07}
Y. Nishiyama, Phys. Rev. E {\bf 75}, 011106 (2007).



%quantum LP
\bibitem{Dutta97}
A. Dutta, B. K. Chakrabarti, and J. K. Bhattacharjee,
Phys. Rev. B {\bf 55}, 5619 (1997).
\bibitem{Dutta98}
A. Dutta, J. K. Bhattacharjee, and B. K. Chakrabarti,
Eur. Phys. J. B {\bf 3}, 97 (1998).

% 3d biaxial nnni
\bibitem{Pal90}
B. Pal and S. Dasgupta,
Z. Phys. B {\bf 78}, 489 (1990).


\bibitem{Dawson88}
K.A. Dawson, M.D. Lipkin, and
B. Widom, J. Chem. Phys. {\bf 88}, 5149 (1988).



% experiment
\bibitem{Huang81}
J. S. Huang and M. W. Kim,
Phys. Rev. Lett. {\bf 47}, 1462 (1981).
%m kotlarchyk S-H Chen JS Huang
%pra 28 508 1983
%M Kotlarchyk S-H Chen JS Huang MW Kim
%pra 29 2054 1984
\bibitem{Honorat84}
P. Honorat, D. Roux, and A. M. Bellocq,
J. Phys. (Paris) Lett. {\bf 45}, L-961 (1984).
\bibitem{Aschauer93}
R. Aschauer and D. Beysens,
Phys. Rev. E {\bf 47}, 1850 (1993).
\bibitem{Seto96}
H. Seto, D Schwahn, M Nagao, E Yokoi, S Komura, M Imai, and K Mortensen,
Phys. Rev. E {\bf 54}, 629 (1996).
\bibitem{Schwahn99}
D. Schwahn, K. Mortensen, H. Frielinghaus, and K. Almdal,
Phys. Rev. Lett. {\bf 82}, 5056 (1999).
%Onajiku
%Physica B 276-278 353 2000
\bibitem{Stepanek02}
P. {\v{S}}t{\v{e}}p\'anek, T. L. Morkved, K. Krishnan,
T. P. Lodge, and F. S. Bates,
Physica A {\bf 314}, 411 (2002).


\bibitem{Nishiyama04}Y. Nishiyama, Phys. Rev. E {\bf 70}, 026120 (2004).




% HIstogram MC
\bibitem{Hellmann93}R.K. Hellmann, A.M. Ferrenberg, D.P. Landau, and R.W. Gerling,
{\it Computer Simulation Studies in Condensed-Matter Physics IV},
edited by
D.P. Landau, K.K. Mon, and H.-B. Sch\"uttler
(Springer-Verlag, Berlin, 1993).
% glass
\bibitem{Shore91}J.D. Shore and J.P. Sethna, Phys. Rev. B {\bf 43}, 3782 (1991).





\bibitem{Roomany80}H.H. Roomany and H.W. Wyld, 
Phys. Rev. D {\bf 21}, 3341 (1980).

%
\bibitem{Deng03}Y. Deng and H.W.J. Bl\"ote,
Phys. Rev. E {\bf 68}, 036125 (2003).

%
\bibitem{Binder81}
K. Binder, Z. Phys. B: Condens. Matter {\bf 43}, 119 (1981).



\end{thebibliography}
\end{document}